\documentclass[runningheads]{llncs}

\usepackage{graphicx}
\usepackage{tikz}
\usepackage{amsmath}
\usepackage{enumitem}   
\usepackage{afterpage}
\usepackage{booktabs} 
\usepackage{array}
\usepackage{hyperref}

\begin{document}
\pagestyle{plain}

\title{A Large-Scale Analysis of Attacker Activity in Compromised Enterprise Accounts}

\author{Neil Shah\inst{1,2} \and
Grant Ho\inst{1,2} \and
Marco Schweighauser\inst{2} \and
M.H. Afifi\inst{2} \and
Asaf Cidon\inst{3} \and
David Wagner\inst{1}}

\authorrunning{N. Shah et al.}

\institute{UC Berkeley \and
Barracuda Networks \and
Columbia University}

\maketitle              

\begin{abstract}
We present a large-scale characterization of attacker activity across 111 real-world enterprise organizations. 
We develop a novel forensic technique for distinguishing between attacker activity and benign activity in compromised enterprise accounts that yields few false positives and enables us to perform fine-grained analysis of attacker behavior.
Applying our methods to a set of 159 compromised enterprise accounts, we quantify the duration of time attackers are active in accounts and
examine thematic patterns in how attackers access and leverage these hijacked accounts. 
We find that attackers frequently dwell in accounts for multiple days to weeks, suggesting that delayed (non-real-time) detection can still provide significant value.
Based on an analysis of the attackers' timing patterns, 
we observe two distinct modalities in how attackers access compromised accounts, 
which could be explained by the existence of a specialized market for hijacked enterprise accounts: 
where one class of attackers focuses on compromising and selling account access to another class of attackers who exploit the access such hijacked accounts provide.
Ultimately, our analysis sheds light on the state of enterprise account hijacking and highlights fruitful directions for a broader space of detection methods,
ranging from new features that home in on malicious account behavior to the development of non-real-time detection methods
that leverage malicious activity after an attack's initial point of compromise to more accurately identify attacks.

\end{abstract}

\section{Introduction}
\label{sec:introduction}
With the growth of cloud-backed services and applications, ranging from email and document storage to business operations such as sales negotiations and time sheet tracking, modern enterprise accounts provide a wealth of access to sensitive data and functionality.
As a result, attackers have increasingly focused on compromising enterprise cloud accounts through attacks such as phishing. For example, several government agencies have issued advisories and reports warning that phishing represents ``the most devastating attacks by the most sophisticated attackers'' and detailing the billions of dollars in financial harmed caused by enterprise phishing and account compromise~\cite{eac-fbi-report,homelandsecurity}. Not limited to financial gain, attackers have also compromised enterprise cloud accounts for personal and political motives, such as in the 2016 US presidential election, when nation-state adversaries dumped a host of internal emails from high-profile figures involved with Hillary Clinton's presidential campaign and the Democratic National Committee~\cite{presidentialEmailLeaks}.

Given the growing importance of online accounts and credentials, a large body of existing work has focused on building mechanisms to defend against attacks through better credential hygiene, detecting phishing attacks, and stronger user authentication~\cite{BECguard,login_challenges,gascon2018reading,ho2019detecting,ho2017detecting,stringhini2015ain,thomas2019protecting}.
Despite these advances, account hijacking, the compromise and malicious use of cloud accounts, remains a widespread and costly problem~\cite{fbi_2020}.
Although prior work has characterized what attackers do with a hijacked account, \cite{bursztein2014handcrafted,onaolapo2016happens,thomas:databreaches},
existing work focuses heavily on compromised personal email accounts. While these insights are useful, it remains unclear how well they generalize to compromised \emph{enterprise} accounts and whether attacks on enterprise accounts have different characteristics. 
Unlike personal accounts, enterprise accounts often have access to a wealth of sensitive business data, and an attacker who compromises one enterprise account can use the identities of the compromised account to launch additional attacks on other employees, expanding their access to other data and assets within the enterprise.

To close this knowledge gap and identify additional avenues for defending enterprise accounts,
we conduct a large-scale analysis of attacker activity within compromised enterprise accounts.
We analyze a historical dataset of nearly 160 real-world compromise accounts from over 100 organizations that have been confirmed as compromised by both a commercial security product (Barracuda Networks) and by the organization's IT or security team.
First, given a compromised account, we develop a method that allows us to identify what account actions correspond to activity by the attacker versus the benign user.Evaluating our approach on a random sample of enterprise accounts, we find that our forensic technique yields a false positive rate of 11\% and a precision of 94\%.

Using this method for fine-grained attacker behavior analysis, we find that for over one-third of the hijacked accounts in our dataset, the attacker's activity occurs across a week or more.
This extended dwell time suggests that there is value in developing (non-real-time) detection techniques that analyze account behavior over longer time horizons to more accurately identify compromise and mitigate an attack before its completed execution.
Additionally, based on a deeper analysis of the access timing patterns within these accounts,
we identify two different modes in the way attackers utilize hijacked accounts.
In particular, the access patterns of hijacked accounts with long durations of attacker activity could reflect the existence of a specialized market of account compromise,
where one set of attackers focuses on compromising enterprise accounts and subsequently sells account access to another set of attackers who focus on utilizing the hijacked account.

Finally, examining the kinds of data and applications that attackers access via these enterprise accounts,
we find that most attackers in our dataset do not access many applications outside of email,
which suggests that either many enterprise cloud accounts do not have access to interesting data and functionality outside of email, 
or that attackers have yet to adapt to and exploit these additional resources.

\section{Data}
\label{sec:data}
Our work starts with a historical dataset consisting of 989 compromised enterprise accounts from 120 real-world organizations. We rely on two pieces of information for ground-truth labeling.
First, all of these organizations use a commercial anti-fraud service (Barracuda Sentinel) for preventing phishing and account takeover attacks \cite{barracuda:sentinel,BECguard}.
For each of the compromised accounts in our dataset,
Barracuda's detectors flagged at least one event (e.g., a user login or a user-sent email) as malicious.
Additionally, all of these compromised instances were verified by their organization's IT or security team.
For the remainder of the paper, we will use the terms \textit{compromised enterprise account}, \textit{compromised user}, \textit{compromised account}, and \textit{account} interchangeably.

\begin{figure}[t]
\begin{center}
\includegraphics[width=\textwidth]{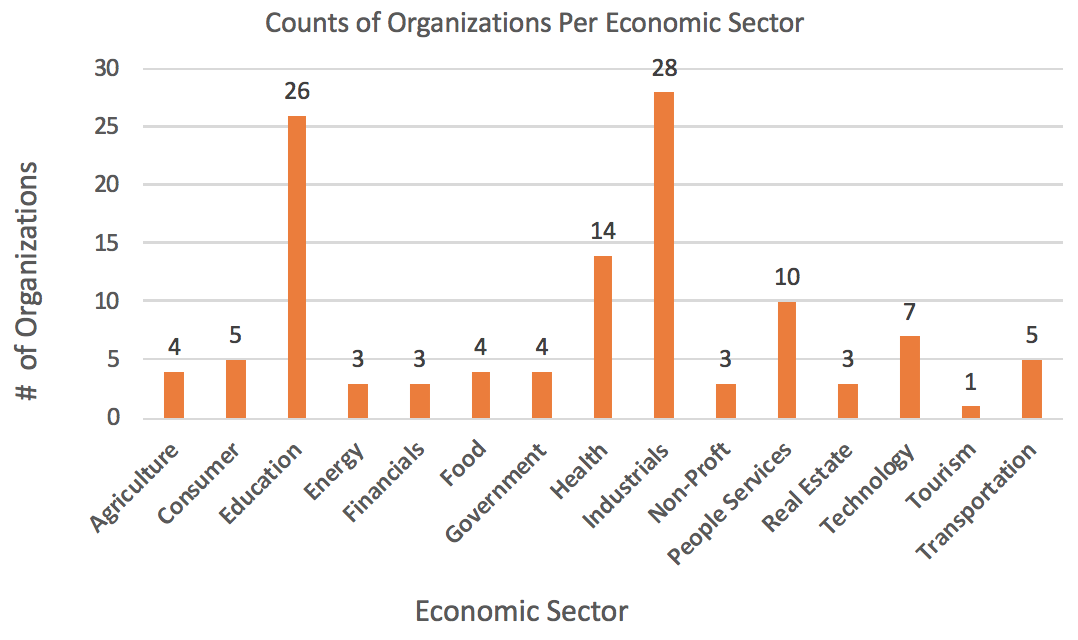}
\end{center}
\caption{\label{fig:economic_sector} Categorization of the 120 organizations in our dataset across various economic sectors.}
\end{figure}

Figure~\ref{fig:economic_sector} shows the distribution of the 120 organizations by economic sector. A majority of these organizations belong to the industrials, education, health, and people services economic sectors, with counts of 28, 26, 14, and 10 respectively. These four sectors represent 65\% of the set of organizations in our dataset.  

\subsection{Schema and Data Sources}
\label{sec:schema}
Our dataset consists of Microsoft Office 365 cloud audit log events~\cite{office365:properties,office365:auditschema}
for all of the compromised accounts we study.
Each time a user logs into their account, accesses an Office 365 application (e.g., Outlook, Sharepoint, and Excel), or performs an account operation (e.g., a password reset), Office 365 records an audit log event.
Across the 989 compromised accounts, our dataset consists of 927,822 audit log events from August 1, 2019 -- January 27, 2020, where each account was marked as compromised during that time window. At a high level, each audit event in our logs includes the following key fields:
\begin{itemize}
  \itemsep0em 
  \item \texttt{Id} - Unique identifier for an audit event
  \item \texttt{UserId} - Email of user who performed the operation
  \item \texttt{UserAgent} - Identifier string of device that performed the operation
  \item \texttt{ClientIp} - IP address of the device that performed the operation
  \item \texttt{Operation} - Operation performed by the user
  \item \texttt{ApplicationId} - Id of Office 365 application acted upon
\end{itemize}
The \texttt{Operation} field indicates which cloud operation the user performed, such as a successful user login or a password reset.
Note that based on the way Office 365 generates these audit events, only events that reflect an account login or application access contain values for \texttt{UserAgent} and \texttt{ClientIp}; audit events for other operations (such as a password reset event) don't contain user agent or IP address information. Throughout the paper, we will refer to the audit events that are login or application accesses as ``application login events" or ``login events". We also augment the information above by using MaxMind~\cite{maxmind} to identify the country and country subdivision (e.g., state or province) of each available Client IP address. 

Additionally, we draw upon several other data sources to help evaluate our technique for distinguishing between benign versus attacker activity within a compromised account (Appendix \ref{sec:rulesevaluation}):
the raw emails sent by users in our dataset and any audit events, emails~\cite{office365:emailmessages}, and inbox forwarding rules flagged by Barracuda's machine learning detectors for the compromised users.

As we discuss in Section~\ref{sec:ruleset}, in order to prevent a large batch of compromised accounts from a single attacker or organization from skewing our analysis results, we de-duplicate this set of 989 compromised accounts to a final dataset of 159 compromised accounts across 111 organizations.

\subsection{Ethics}
\label{sec:ethics}
This work reflects a collaboration between researchers at UC Berkeley, Columbia University, and a large security company, Barracuda Networks. The set of organizations included in our dataset are customers of Barracuda, and is secured using standard industry best practices.

Due to the confidential nature of account data, only authorized employees of Barracuda Networks accessed the data, and no sensitive data was released to anyone outside of Barracuda Networks. 
Our project received approval from Barracuda Networks, and strong security controls were implemented to ensure confidentiality and limited scope of access.


\section{Detecting Attacker Activity}
\label{sec:detection}
Compromised accounts contain a mix of activity, such as application accesses (logins), from both the true user and an attacker.
In order to accurately analyze attacker usage of hijacked accounts, we developed a ruleset, based on well-known anomaly detection ideas~\cite{auth0,robertson:anomalydetection}, for identifying which audit events correspond to activity from the attacker versus a benign user.

Throughout this section, when describing the components of our rule set, we use the name Bob to refer to a generic compromised user from our dataset. Our rule set first builds a historical profile for Bob that represents the typical locations and user agent strings that he uses to log into his account.
We then use this profile to classify future login events as either attacker-related or benign by identifying actions that deviate from the historical profile.
Our rule set is not guaranteed to find every attack, nor does it guarantee robustness against motivated attackers trying to evade detection. However our rule set is still relatively comprehensive and generates few false positives. 

\subsection{Historical User Profile and Features}
\label{sec:historicalprofilefeatures}
\subsubsection{Historical User Profile}
Conceptually, a user's historical profile reflects the typical activity (operations, login provenance, etc.) that the user makes under benign circumstances.
To construct this profile, we assume that historical login events that occurred significantly (one month) before any known compromise activity reflect benign behavior by the true user. For each compromised user (Bob), we find the earliest time, $t$, that any of Barracuda's detectors flagged the account as compromised.
To create Bob's historical user profile, we first retrieve a set of historical login events from the time window of 2 months prior to $t$ until 1 month prior to $t$ (i.e., one-month of historical data). From this historical dataset, we construct a historical profile for Bob that consists of 3 sets of values: the set of country subdivisions (states or provinces) that he logged in from during that time period, the set of countries he has logged in from, and the set of user agents that he has logged in with.

\subsubsection{Features}
Given a recent event, $e$, that we wish to classify as malicious or benign activity, we extract 2 features based on a user's historical profile. First, we extract a numerical geolocation feature by comparing the geolocation of $e$'s IP address to the set of geolocations in the user's historical profile: 
\begin{enumerate}[label=(\alph*)]
    \item If $e$ represents a login from a country that was never seen in Bob's historical user profile, then assign $e$'s geolocation feature value a \textbf{2} (most suspicious).
    \item Otherwise, if $e$ represent a login from a country subdivision not found in Bob's historical user profile, then assign $e$'s geolocation feature value a \textbf{1} (medium suspicion). 
    \item Otherwise, assign $e$'s geolocation feature value a \textbf{0} (least suspicious).
\end{enumerate}

We also extract a user agent feature that captures the suspiciousness of the user agent of $e$. All user agents are normalized in a pre-processing step: the version number is removed and only the device and model identifiers are retained, so a user agent string such as \texttt{iPhone9C4/1706.56}
is normalized to \texttt{iPhone9C4}. Thus, \texttt{iPhone9C4/1706.56} and \texttt{iPhone9C4/1708.57} yield the same normalized user agent. The user agent feature is then defined as follows:

\begin{enumerate}[label=(\alph*)]
    \item If $e$'s normalized user agent does not match any of the normalized user agents in Bob's historical user profile, then assign $e$'s user agent feature value a \textbf{1} (most suspicious).
    \item Otherwise, assign $e$'s user agent feature value a \textbf{0} (least suspicious). 
\end{enumerate}

\subsection{Classification Rule Set}
\label{sec:ruleset}
In order to identify the set of attacker actions within a compromised account, we start by selecting the first known compromise event that Barracuda's detectors marked as malicious and that was confirmed by the organization's IT team.
Next, we compute a user's historical profile as described above and use it to extract features for every login event in a two-month window centered around this first confirmed compromise event (i.e., all login events in the month preceding this initial compromise time as well as all login events in the one month following the initial compromise time).
We then apply the following set of rules to classify the login events in this ``recent'' two-month window as attacker activity or not.
Below, we present a high-level sketch of our rule set and discuss assumptions made in the development of our rules. We defer further details and evaluation results to Section~\ref{sec:rulesassumptions} and Appendix~\ref{sec:rulesevaluation}.

\subsubsection{Rules} 
For a compromised account (Bob), each recent event contains a geolocation feature, denoted as \textbf{geo}, and a user agent feature, denoted as \textbf{ua}, as described above.
Given these features, we mark an event as malicious or benign based on the following rule set:
\begin{verbatim}
    if geo == 2 
        mark e as malicious (attacker related)
    else if (geo == 1) and (ua == 1)
        mark e as malicious
    else 
        mark e as benign
\end{verbatim}

\subsubsection{Intuition and Assumptions}
\label{sec:rulesassumptions}

The geolocation and user agent features quantify the suspiciousness of a new login event in relation to a user's historical profile. We assume that the historical login events for each user do not contain attacker activity;
i.e., that the attacker has not operated within the account for over one month prior to detection.
However, it is possible that some of the events shortly preceding the initial confirmed compromise could be attacker related.
Thus, we conservatively analyze one month's worth of events preceding a user's first confirmed compromise event to more comprehensively capture and analyze the full timeline of an attack.

Our rule set also assumes that it is less common for users to travel to another country than to another state or province within their home country. Although traveling abroad is common in some industries, we assume that most employees travel more frequently to another state or region within their residential country rather than to an entirely different country.
As a result, if a login event contains an IP address mapped to a country that was never seen before in a user's historical login events, the event in question is marked as an attacker event. For travel within the same country, the country subdivision and user agent need to be new for a login event to be marked as an attacker event.

\subsubsection{Applying Rule Set to Compromised Users} \label{sec:applying} For each user Bob, we classify all login events from one month prior to $t$ to $t$ using a historical user profile based on events from two months prior to $t$ to one month prior to $t$. Then, we classify all events from $t$ to one month after $t$ using a historical user profile based on events from two months prior to $t$ to one month prior to $t$, and all events from one month prior to $t$ to $t$ that were classified as benign. Thus we update the historical user profile for each user after classifying the first month of login events \cite{aws:retraining}. Malekian et al. also describes a similar approach \cite{malekian:adaptivehistoricalprofile} where the historical profile is updated to reflect new patterns in user behaviors in e-commerce for the purposes of detecting online user profile-based fraud. Therefore, the last month of Bob's events are classified using an updated historical user profile that incorporates benign activity from his previous month of login events.

After applying this rule set to the 989 compromised accounts in our dataset, we identified 653 accounts (across 111 organizations) that contained at least one attack event.
276 of the 989 compromised users didn't have any historical login events due to the fact that these users' enterprises registered with Barracuda as a customer after the start of our study period, and we did not have login events from before then. As a result, our rule set couldn't be applied to these users. Of the remaining 713 users that had historical login events, 653 had at least one attacker event that our rule set classified.

We also found that 68\% of the 653 compromised accounts belonged to only 6 organizations. We do not know what accounts for this skewed distribution, but it is possible that one or a few attackers specifically targeted those 6 organizations.
Therefore, to ensure that our analysis results in Section~\ref{sec:findings} are not biased by a few attackers that compromised many accounts, we randomly sampled a subset of compromised accounts from each of the 111 organizations, resulting in a dataset of 159 compromised accounts that we use for our analysis in Section~\ref{sec:findings}. Appendix \ref{sec:rulesetsampling} contains more details about our sampling procedure, as well as a detailed breakdown of the 653 compromised users across the 111 organizations.

In order to evaluate the accuracy of our rule set at labeling an event as malicious or not, we randomly sampled a set of 20 compromised accounts and manually labeled each event based on the analysis procedure described in Appendix \ref{sec:rulesevaluation}.
Our evaluation suggests that our rule set has a false positive rate of 11\% and precision of 94\%.

\subsubsection{Limitations of Rule Set and Attacker Evasion} Although our rule set has relatively high precision, we acknowledge some limitations that exist with our rules and features. Given the construction of our rule set, if a motivated attacker logs in from a state that the typical user has logged in from or with a device and model that matches that of the typical user, the attacker would successfully evade our rule set. 

We did observe evidence of attackers trying to ``blend in'' with benign characteristics of some users, potentially to evade detection.
For the 60 compromised enterprise accounts mentioned above in Section \ref{sec:applying} in which our rule set classified no events as attacker-related, we took a random sample of 10 accounts and performed deeper analysis of the events that our rule set classified as benign.
For 6 of the 10 accounts, we found that attackers only logged in from locations close in proximity to those logged in by the true user of the account (within the same states as the typical user). 
The geolocations appeared normal and since all 10 of these accounts were flagged by Barracuda's detectors, this is evidence of likely evasive behavior.
This potentially evasive behavior parallels a result from Onaolapo et al.~\cite{onaolapo2016happens}, where they found that attackers deliberately choose their geolocations to match or come close to ones used by the true user in an effort to evade detectors that look for anomalous geolocations. 

For the remaining 4 accounts, we see a combination of logins from close geolocations to ones used by the true user and further geolocations (e.g. different province), but it is unclear if the logins from different provinces originate from the attacker or not given the similar user agent strings (same device and model) that are present in the event logs. This could be potential evidence for user agent masquerading, but additional work would be needed to explore this idea further.

\section{Characterizing Attacker Behavior}
\label{sec:findings}

In this section, we conduct an analysis of attacker behavior across our dataset of 159 compromised users belonging to a total of 111 organizations.
Our analysis reveals three interesting aspects of modern enterprise account hijacking.
First, we find that for a substantial number of accounts (51\%), malicious events occur over multiple days.
From a defensive standpoint, this suggests that while real-time detection is ideal,
detectors that identify attacks in a delayed manner (e.g., as a result of using features based on a longer timeframe of activity) might still enable an organization to thwart an attack from achieving its goal.
Second, we observe evidence that at least two distinct modes of enterprise account compromise exist.
In particular, we estimate that 50\% of enterprise accounts are compromised by attackers who directly leverage the information and access provided by the hijacked account.
In contrast, for roughly one-third of the the compromised accounts in our dataset, the attackers' access patterns suggest a compromise strategy where one set of attackers compromised the account and then sold access to the account (i.e., its credentials) to another set of actors who ultimately leveraged the account for malicious purposes (e.g., by sending spam or phishing emails).
Finally, we find that attackers who compromise enterprise accounts primarily use the accounts for accessing email-related information and functionality; 78\% of the hijacked accounts only accessed email applications across all their attacker events.
Given that attackers did not access other applications (such as SharePoint or other corporate cloud applications), this suggests that a number of real-world attackers have not yet investigated or found value in accessing other data and functionality provided by these enterprise accounts, outside of email.

 \subsection{Duration of Attacker Activity and Damage Prevention}
\label{sec:duration}
In this section, we estimate the length of time attackers are active in enterprise accounts. 
Our results suggest that in many cases, attackers spend multiple days exploiting the information and functionality within enterprise accounts. This suggests that even if a detector doesn't operate in a real-time fashion, it can still prevent attackers from inflicting significant damage.

\subsubsection{Duration of Attacker Activity}
Given our dataset of 159 compromised users and their respective login events, we cannot definitively determine how long an attacker compromised the account for. However, we can estimate a reasonable lower bound on the length of time an attacker is active within an account (i.e. logging in and accessing the account).
\begin{figure}
\begin{center}
\includegraphics[width=\textwidth]{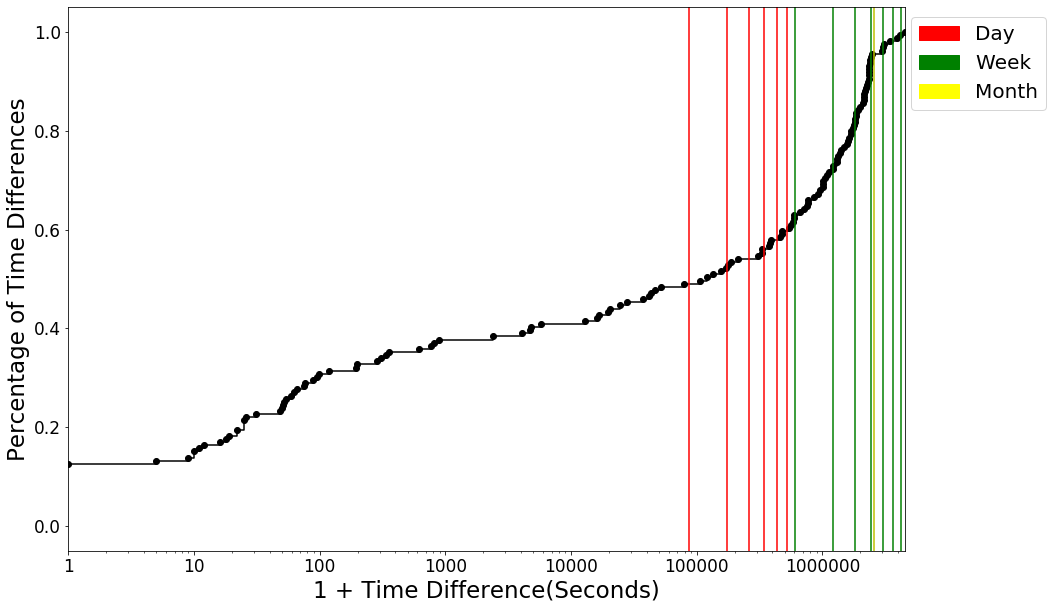}
\end{center}
\caption{\label{fig:firstlastlogin} The distribution of time (seconds) between the first and last attacker login event across the compromised accounts in our dataset.}
\end{figure}
For each user, we computed the difference (in seconds) between the time of the earliest attacker login event and the time of the last attacker login.
As seen in Figure \ref{fig:firstlastlogin}, across all 159 compromised enterprise accounts, attackers appear to use and maintain access to many enterprise accounts for long periods of time.
In almost 51\% of the enterprise accounts within our dataset (81 out of 159), attackers are active for at least 1 day and in 37\% of accounts, attackers are active for at least 1 week.
As a result, while it's important to detect attacks in real-time, detection can still provide significant value even if it occurs after the initial compromise.

As an example of where non-real-time detection can still mitigate significant harm, we analyzed accounts that sent at least one malicious email flagged by one of Barracuda's detectors during the two month ``attack window'' that we applied our rule set on.
Across the 11 corresponding accounts, 7 out of the 11 accounts (63\%) exhibited a 3 day gap between the first malicious login event identified by our rule set and the first phishing email sent by the account (Appendix \ref{sec:damageprevention} shows the full interarrival distribution for all 11 accounts).
In these instances, with a long gap between the initial compromise and the first phishing email,
a detector that uses more computationally expensive features or detection methods, which might not be feasible to run in real-time, could prevent a significant portion of the attack activity. In the absence of such a detector, even providing manual tools for organizations to investigate whether the compromised account affected additional ones may prove beneficial.
  
\subsection{Attacker Account Access Patterns}
\label{sec:multipleattackers}
In this section, we explore the different modes in which attackers access these hijacked accounts.
We estimate that in 50\% of our sample of enterprise accounts, a single attacker conducts both the compromise and utilization of the account.
However, for many of the remaining accounts, both the timing and application access patterns suggest that potentially two or more attackers compromise and access the hijacked account.
This access pattern would be consistent with the existence of a specialized market for compromised enterprise accounts, where one set of attackers conducts the compromise and another attacker buys access to the compromised account and obtains value from the account (e.g., by accessing sensitive information or sending spam or phishing emails). 

\subsubsection{End-to-End Attackers} Revisiting our findings from Section \ref{sec:duration}, we found that 81 out of 159 enterprise accounts (51\%) are compromised for at least 1 day, suggesting that there are largely two main segments of compromised enterprise accounts; those that are compromised for less than a day and the remaining that appear to be compromised for a day or more. Given this preliminary result, we aim to investigate the relationship between duration of attacker activity and the economy and existence of various modes of attackers operating in the enterprise account space.

We start by investigating whether enterprise accounts are generally accessed regularly by attackers or in isolated pockets of time during the compromise lifecycle. For each of the 159 compromised enterprise accounts, we compute the interarrival time (absolute time difference) between every pair of successive attack events sorted in time. We then take the max interarrival time for each user, which represents the longest time gap between any two successive attacker accesses within an account. 
\begin{figure}
\begin{center}
\includegraphics[width=\textwidth]{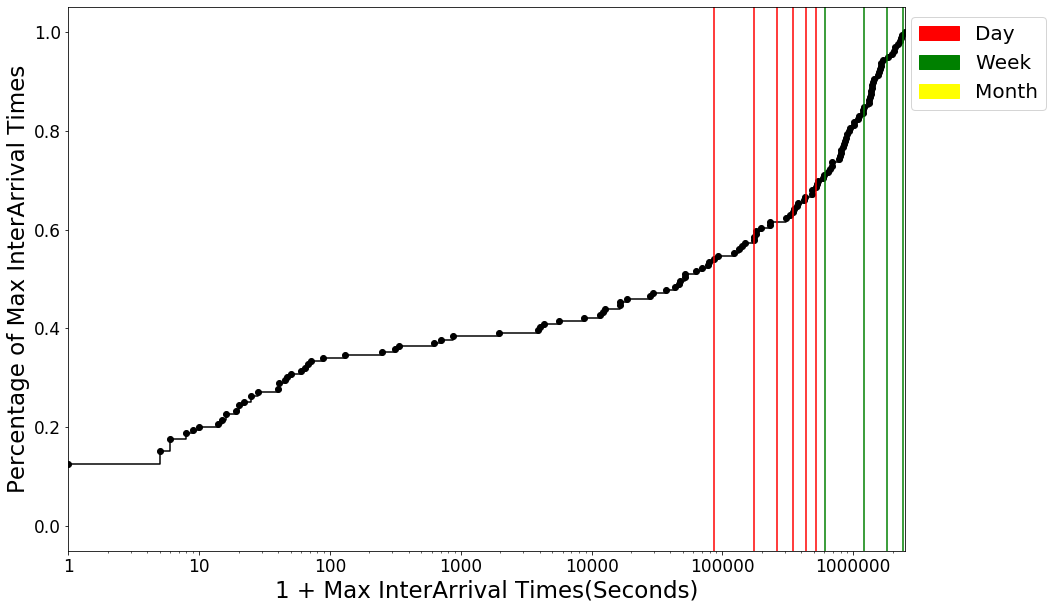}
\end{center}
\caption{\label{fig:maxinterarrival} Distribution of the maximum attacker interarrival times (seconds) for all 159 enterprise accounts. The maximum attacker interarrival corresponds to the longest time gap between two consecutive attack events.}
\end{figure}
From Figure \ref{fig:maxinterarrival}, which shows a CDF of the max attacker interarrival times in seconds for all 159 compromised enterprise accounts, we can see that at around the 1 day mark (first red line from the left), the inflection and trend of the CDF start to change. In 53\% of compromised enterprise accounts, the largest time gap between successive attacker accesses is less than 1 day, while the remaining 47\% of compromised enterprise accounts (74 out of 159) have 1 or more days as their longest time gap. 

A short attack life cycle (i.e., less than 1 day) seems to reflect an end-to-end compromise approach: where a single actor compromises an account and also leverages its access for further malicious actions.
In our dataset, 78 out of the 159 enterprise accounts (50\%) fall within this category.
Due to the small time gaps between successive attacker events and relatively small durations of time attackers are active, these 78 accounts are likely compromised by a single set of attackers that both perform the compromise and use the accounts for a short period of time; it is also possible that some of these cases reflect compromise within an organization that rapidly identified and responded to the compromise incident.

\subsubsection{Segmented Account Access}
As seen in Figure \ref{fig:maxinterarrival}, 53\% of enterprise accounts (74 out of 159) experienced a maximum of 1 or more days between successive attacker events.
One possible explanation of the large time gap is that the initial set of attackers that compromised these accounts sold them to another set of attackers; hence, the time gaps represent the time needed for the transaction to complete.
Exploring this theory, we compared attacker events before and after the max attacker interarrival time in these 74 accounts on the basis of geolocation, user agent, and internet service providers (ISPs). 
If the two periods of activity have significant differences across these three attributes, then that suggests that the two different activity windows could reflect access by two different sets of attackers.

To quantify the similarity of the two sets of attributes before and after the max interarrival time, we use the \textit{Jaccard Similarity Coefficient}. Given two sets of data $A$ and $B$, the Jaccard Similarity Coefficient relates the number of elements in the set intersection of $A$ and $B$ to the number of elements in the set union of $A$ and $B$. It has been widely used in many fields \cite{securempc:jaccard,testcase:selection,jaccard:keywordsimilarity,wu:extractingtopics} such as keyword similarity matching in search engines to test case selection for industrial software systems.

For each of the 74 compromised enterprise accounts, we gather two sets of country subdivisions mapped to attacker events before and after the max attacker interarrival time respectively. Similarly, we gather two sets of user agents and two sets of ISPs in the same manner. We then compute 3 Jaccard similarity coefficients for geolocation, user agent, and ISP respectively. 
\begin{figure}[t]
\begin{center}
\includegraphics[width=\textwidth]{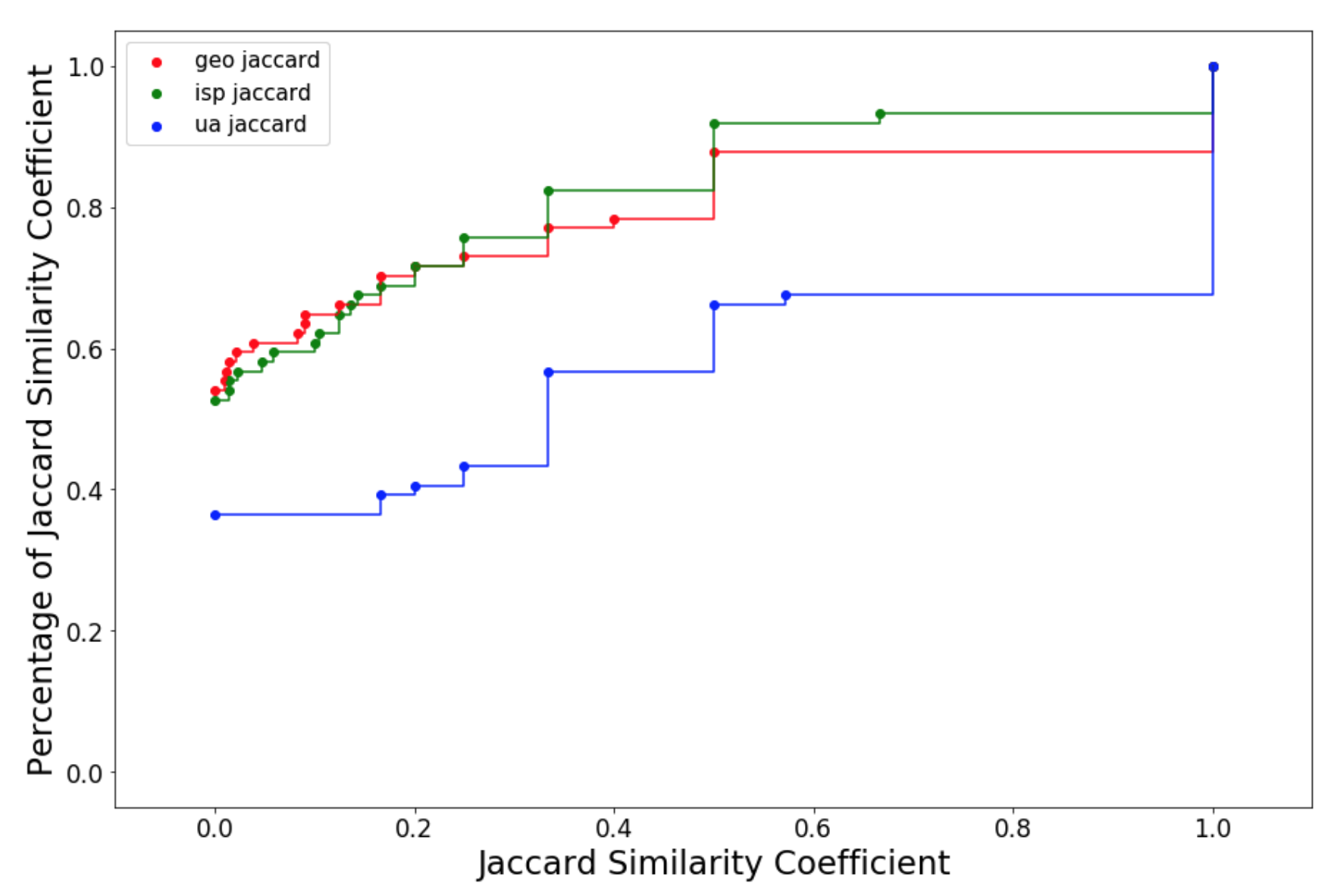}
\end{center}
\caption{\label{fig:jaccard} Distribution of the Jaccard Similarity between Geolocation, User Agent, and ISP usage across the two time-separated attacker access periods; 74 compromised enterprise accounts had long time gaps (max interarrival times) between attacker access events.}
\end{figure}
In Figure \ref{fig:jaccard}, most of the enterprise accounts have low Jaccard similarity coefficients for geolocation and ISP; one reason the user agent curve follows a different pattern is because of the normalization we performed, where we treat user agent strings with different device versions as the ``same'' underlying user agent. 
50 of the enterprise accounts (around 70\% of  74) had Jaccard similarity coefficients of 0.3 or less for geolocation and ISP, indicating that the sets of country subdivisions and ISPs before and after the large time gaps in these accounts were substantially different. 

We also show in Appendix \ref{sec:stability} that if attackers are using proxy services for obtaining IP addresses, in 85\% of the 74 compromised enterprise accounts, these services are fairly stable; hence, the low geolocation Jaccard similarity coefficients are not a result of attackers using unstable anonymized IP proxies or even Tor.

Given the large time gaps between successive attacker events and low similarity of the access attributes between these attack periods,
we believe that 50 of the 159 enterprise accounts (31\%) reflect the exploitation of a hijacked account by multiple attackers.
For example, these hijacked accounts could reflect compromise that results from a specialized economy, where one set of attackers compromise the accounts and sell the account credentials to another set of attackers that specialize in monetizing or utilizing the information and access provided by the enterprise account.

In terms of understanding the potential damage inflicted by the two sets of attackers, we found that in 30 of the 50 accounts (60\%), the second set of attackers that utilize the accounts access Office 365 applications at a higher rate than the first set of attackers. This further shows the importance of early mitigation in compromised enterprise accounts and that non-real-time detectors should be designed to monitor continuous activity in order to prevent future damage after an account is sold. Details of our analysis are shown in Appendix \ref{sec:activitydiff}. 

Overall, in this section, we identified two distinct patterns of compromise and access behavior that reflect attacker behavior across 81\% of enterprise accounts. For many of these accounts, significant differences between the attacker's login location and access patterns suggest that modern enterprise account exploitation consists of two phases conducted by separate types of attackers: those who compromise the account and those who leverage the hijacked account's information and functionality.

\subsection{Uses of Compromised Enterprise Accounts}
\label{sec:applicationaccesses}
In this section, we explore how attackers use enterprise accounts.
In our dataset, attackers do not appear to establish additional access footholds into the account: they rarely change account passwords and never grant new OAuth access. In addition, within the Office 365 ecosystem, we find that attackers are not very interested in many cloud applications outside of email; 78\% of the enterprise accounts only accessed email applications through attack events.

\subsubsection{Other Operations Performed During Attacker Window} 
As we discussed in Section \ref{sec:data}, every audit event has an \texttt{Operation} field that specifies the action that was taken. The operations we are most interested in learning if attackers perform are ones that affect a user's ability to access their account; namely, operations such as ``Change user password" and ``Add OAuth". The operation ``Change user password" enables the user to change the password to their account, while the ``Add OAuth" operation enables a user to grant applications access to certain data within their account. Since our rule set only classifies login events due to the non-empty IP and user agent fields, we gather all ``Change user password" and ``Add OAuth" audit events that are close in time to each account's attack events. 

We find that only 2 out of 159 compromised enterprise accounts (2\%) had at least one ``Change user password" operation performed close in time to attacker activity. Looking deeper into the 2 accounts, we see the presence of more attacker activity after the change password operations were performed, indicating that these operations were performed by the attacker themselves. None of the 159 accounts had a single ``Add OAuth" operation performed during the time period of attacker activity. Taken together, these findings suggest that attackers are not interested in changing a user’s password or adding OAuth to a user’s account, as this might reveal to the user that their account has been compromised and limit the amount of time the attacker can operate in the account. As a result, a ``Change user password'' event or ``Add OAuth'' event are likely not good features for detectors, as they are rarely found performed by an attacker.

\subsubsection{Unusual Application Accesses by Attackers} 
We now aim to understand if there are specific Office 365 applications outside of frequently accessed email applications, such as Microsoft Exchange and Microsoft Outlook, that attackers access but the true users of the accounts don't access.

There were a total of 21 non email-related Office 365 applications that were accessed by at least one of the 159 accounts.
\begin{figure}[t]
\begin{center}
\includegraphics[width=\textwidth]{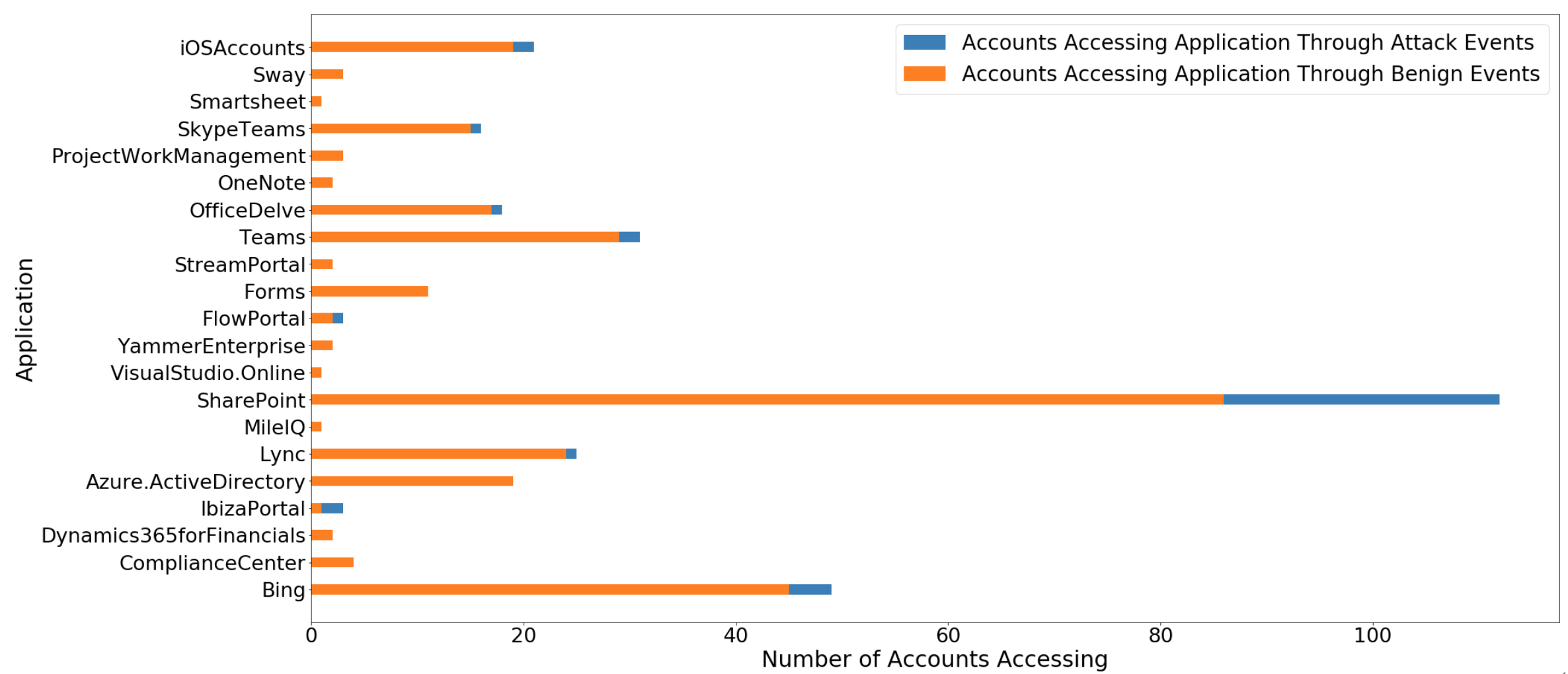}
\end{center}
\caption{\label{fig:newapplications} Bar chart comparing number of accounts accessing each of the 21 non-email applications via only attacker-labeled events and number of accounts accessing non-email applications via only benign events from August 1, 2019 -- January 27, 2020.}
\end{figure}
For each of the 21 non-email applications, we determined the number of accounts that only accessed the application through their attack events and the number of accounts that only accessed the application through their benign events. The results for each of the 21 non-email applications are shown in the stacked bar chart in Figure \ref{fig:newapplications}. Surprisingly, other than Ibiza Portal, none of the remaining 20 applications had the characteristic of more accounts accessing it only through attack events than number of accounts accessing it through benign events. 3 accounts accessed Ibiza Portal only through attack events, while only one account accessed it solely through benign events. Ibiza Portal, or Microsoft Azure portal, \cite{microsoft:azureportal} enables users to build and monitor their enterprise's web and cloud applications in a simple, unified place; therefore, it might allow an attacker to view confidential data within an enterprise's applications, but retrieving that data may take longer compared to other file-sharing applications, such as Microsoft SharePoint or Microsoft Forms. In addition, Microsoft Azure Portal is very rarely accessed by true users of enterprise accounts (only one account ever accessed Microsoft Azure Portal during their benign events). Therefore, based on our dataset of compromised enterprise accounts, it does not appear that attackers are accessing cloud-based applications that typical users don't access within the Office 365 ecosystem. Therefore, in the current state, building features for detectors around atypical accesses to cloud-based applications may not aid much in detecting attacker activity post-compromise. Future work would involve exploring additional cloud-based applications outside of Office 365.

\subsubsection{Applications that Attackers Commonly Use} 
In this section, we aim to understand the types of cloud applications that attackers exploit in enterprise accounts, regardless of how common the application is for enterprises to use.

Most attackers favor email-related applications. We found that in 98\% of compromised enterprise accounts (156 out of 159), attackers accessed at least one email-related Office 365 application. Much of the previous work in understanding compromised personal accounts found that attackers tended to go through user's inboxes and send phishing emails; we now see that at scale, attackers seem to be exhibiting similar behavior in enterprise accounts. We also found that in 78\% of compromised enterprise accounts (124 out of 159), attackers only accessed email-related Office 365 applications. We speculate that this may be because examining a user's inbox is sufficient for attackers who want to learn more about the user and the enterprise the account belongs to. 

In terms of non-email-related Office 365 applications, Microsoft Sharepoint has the highest percentage of accounts that access it through attack events (17\%), with Bing as the second highest percentage at 3\%. A full bar chart showing the percentage of enterprise accounts that access each non-email related Office 365 application through attack events is shown in Appendix \ref{sec:appsfavor}. Given the wide range of Office 365 cloud applications accessible by attackers and the abundance of information these applications harbor, it is surprising that attackers don't access these applications more often. Attackers of enterprise accounts still generally favor email-related applications, such as Microsoft Outlook, which offer a quick and convenient way for an attacker to gain access to contact lists and learn about any confidential and financial information tied to the employee and or enterprise.  

\section{Related Work}
In this section, we highlight previous works that study detection and characterization of compromised accounts. We also draw comparisons between our findings in the space of compromised enterprise accounts and that of previous work.
\subsection{Overview of Previous Work}
\subsubsection{Detection and Forensics} There has been an extensive amount of literature proposing various techniques from machine learning and anomaly detection for detecting phishing attacks in personal and enterprise accounts on a smaller scale \cite{abu:phishingdetection,bergholz:phishingdetection,fette:phishingdetection,hu:detecting} and on a large scale \cite{BECguard,ho2019detecting,ho2017detecting,stringhini2015ain}. In addition, a limited amount of prior work exists on detecting compromised accounts \cite{egele:compa,liu:honeyfiles} through the use of honeypot accounts and personal accounts on social networks. 

Liu et al. in \cite{liu:honeyfiles} monitored the dangers of private file leakage in P2P file-sharing networks through the use of honeyfiles containing forged private information. Their work focused more on the use of honeyfiles instead of account credentials and doesn't study compromised accounts outside of P2P. 

Egele et al. in \cite{egele:compa} developed a system, called COMPA, for detecting compromised personal accounts in social networks. COMPA constructs behavior profiles for each account and evaluates new messages posted by these social networking accounts by comparing features such as time of day and message text to the behavioral profiles. They measured a false positive rate of 4\% on a large-scale dataset from Twitter and Facebook. However, their work only studies how to detect compromised personal accounts and doesn't include enterprise accounts. 

As a result, none of the works in the literature have performed analysis to understand attacker activity in enterprise accounts post-compromise. Our work addresses this gap by presenting a forensic technique that allows an analyst or organization to distinguish between attacker and benign activity in enterprise accounts. 

\subsubsection{Characterization} Although there has been prior work on understanding attacker behavior and patterns within compromised accounts ~\cite{bursztein2014handcrafted,fairbanks:forensics,onaolapo2016happens,thomas:twitter}, most of this research has been primarily focused on understanding the nature of compromised personal accounts; few efforts have been examined the behavior of attackers in compromised enterprise accounts at large scale. 

TimeKeeper, proposed by Fairbanks et al.~\cite{fairbanks:forensics}, explored the characteristics of the file system in honeypot accounts controlled by attackers. Although their work applied forensic techniques to honeypot accounts post-compromise, they operated at small scale and only characterized attacker behavior in relation to file systems on these accounts. 

Onaolapo et al~\cite{onaolapo2016happens} studied attacker behavior in small-scale hijacked Gmail accounts post-compromise and characterized attacker activity based on where the account credentials were leaked. They also devised a taxonomy of attacker activity accessing the Gmail accounts, noting the presence of four attacker types (curious, gold diggers, spammers, and hijackers). However, their work did not examine compromised enterprise accounts and they were only able to monitor certain actions, such as opening an email or creating a draft of an email. 

Bursztein et al.~\cite{bursztein2014handcrafted} examined targeted account compromise through the use of various data sources, such as phishing pages targeting Google users and high-confidence hijacked Google accounts. However, their work focuses on compromised personal accounts and not on enterprise accounts. 

\subsection{Comparing Enterprise versus Personal Account Hijacking}
\subsubsection{Duration of Attacker Activity} Extensive prior works have studied how long attackers remain active within personal accounts, but none have studied this characteristic in compromised enterprise accounts. 
Thomas et al.~\cite{thomas:twitter} studied account hijacking in the context of social media by analyzing over 40 million tweets a day over a ten-month period originating from personal accounts on Twitter. They found that 60\% of Twitter account compromises last a single day and 90\% of compromises lasted fewer than 5 days. However, in our work with compromised enterprise accounts, we find that in 37\% of accounts, attackers maintain their access for 1 or more weeks. 

Onaolapo et al.~\cite{onaolapo2016happens} also found that the vast majority of accesses to their honey accounts lasted a few minutes or less.
However, their work also notes that for about 10\% of accesses by "gold digger" attackers (those that search for sensitive information within an account) and for most accesses by "curious" attackers (those that repeatedly log in to check for new information), attacker activity lasted several days.
These two modalities, of short and long compromise durations, also manifests itself in our results,
where attackers in nearly half of the compromised accounts in our dataset conducted all of their activity within one day, but over one-third of hijacked accounts experienced attacker activity across multiple days or weeks.

\subsubsection{Attacker Account Usage Patterns.}  Onaolapo et al. also devised a taxonomy of attacker activity and categorized four different attacker types (curious, gold diggers, spammers, and hijackers) based on personal honeypot accounts leaked to paste sites, underground forums, and information-stealing malware. Additionally, Onaolapo et al. found that the largest proportion of "gold digger" accesses came from honey accounts leaked on underground forums where credentials are shared among attackers.
In our work, we explore the potential for an economy of compromised enterprise accounts and the different modes in which attackers access these hijacked accounts. We estimate that in 50\% of our sample of enterprise accounts, a single attacker conducts both the compromise and utilization of the account. 
Additionally, we find that roughly one-third of accounts in our dataset appear to be accessed by multiple attackers; one explanation for this could be the existence of a specialized market for compromised enterprise accounts where one attacker conducts the compromise and another attacker likely buys access to the compromised account and obtains value from the account (e.g., by accessing sensitive information or sending spam or phishing emails). Such an economy, where compromised enterprise accounts are also sold in underground forums, would be consistent with the findings in Onaolapo et al.

\subsubsection{Uses of Compromised Accounts.} Much of the prior work in the space of enterprise and personal accounts has studied attacker activity from the perspective of detecting phishing emails. For example, Ho et al. \cite{ho2019detecting} conducted the first large-scale detection of lateral phishing attacks in enterprise accounts. We find that within the space of compromised enterprise accounts, email-related applications still seem to be the most common and desired way attackers obtain information within accounts. This suggests that either many enterprise cloud accounts may not have access to interesting data outside of email or that attackers have yet to exploit these additional sources of information in enterprise accounts. As a result, email remains an important direction of analysis within the space of compromised enterprise accounts.

\section{Summary}
In this work, we presented a large-scale characterization of attacker activity in compromised enterprise accounts. We developed and evaluated an anomaly-based forensic technique for distinguishing between attacker activity and benign activity, enabling us to perform fine-grained analysis of real-world attacker behavior. We found that attackers dwell in enterprise accounts for long periods of time, indicating that in some situations, non-real-time detectors that leverage more computationally expensive approaches and features can still provide significant defensive value.
Based on the timing of attacker behavior, we infer that a specialized market for compromised accounts might exist, with some attackers developing skills specific for stealing credentials and other attackers specializing in how to extract information and value from a hijacked account.
Finally, we find that most attackers in our dataset do not access many applications outside of email, which suggests that attackers have yet to explore the wide-range of information within cloud applications. 

\section{Acknowledgements}
We thank Professor Raluca Ada Popa and the anonymous reviewers for their extremely valuable feedback. This work was supported in part by the Hewlett Foundation through the Center for Long-Term Cybersecurity, an NSF GRFP Fellowship, and by generous gifts from Google and Facebook.

\appendix

\section{Rule Set Extended Details}
\subsection{Extended Details on Applying Rule Set}
\label{sec:rulesetsampling}
\begin{figure}
\begin{center}
\includegraphics[width=\textwidth]{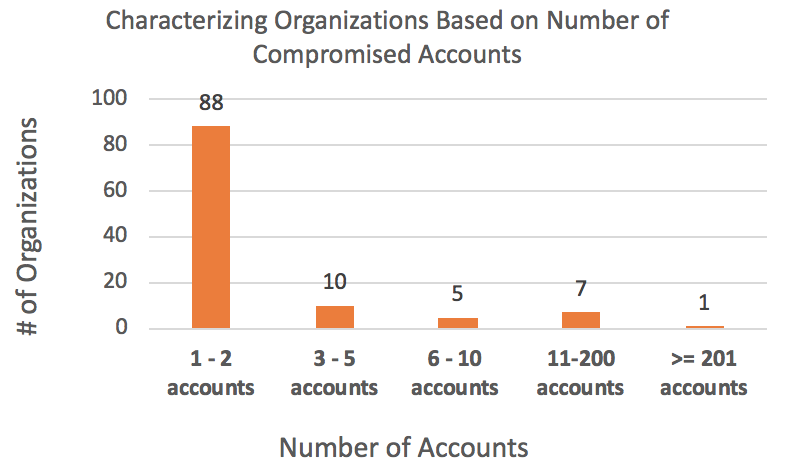}
\end{center}
\caption{\label{fig:orgs_num_users} Categorization of the 111 organizations in our dataset based on number of compromised user accounts.}
\end{figure}
After applying our rule set on the original set of 989 compromised users, we obtained 653 compromised users that had at least one attacker event classified. Across these 653 compromised users, our attacker rule set labeled 17,842 audit events as attacker-related. Figure \ref{fig:orgs_num_users} shows the distribution of compromised accounts among organizations. 98 of the 111 organizations (89\%) had 1--5 compromised users, 12 organizations had 6--200 compromised users, and 1 organization had over 200 compromised users, precisely 206. Moreover, 68\% of the 653 compromised accounts belong to 6 organizations. As a result, it is possible that one or a few attackers specifically targeted those 6 organizations. Therefore, to ensure that we obtain many different attackers and our results are not biased by a few attackers that compromise many accounts, we grouped each of the 653 compromised users by organization and month of earliest attack event flagged by our rule set and randomly selected one compromised user from each group. This resulted in a final sample of 159 compromised users across the 111 organizations.

\subsection{Evaluation}
\label{sec:rulesevaluation}
In this section, we evaluate our attacker rule set. We first describe how we sampled a set of users and established ground truth labels for login events. We then show how well our rule set performs on the sample of users. Overall, our rule set generates few false positives and offers a promising direction for distinguishing between attacker activity and benign activity at the granularity of login events.

\subsubsection{Evaluation Methodology}
\label{sec:evaluationmethodology}
We evaluated our rule set on 159 compromised enterprise accounts that each have attacker events classified by the rule set. To understand how effective our rules are, we randomly sample 20 users and manually evaluate the accuracy of the rule set on these users. For each of the 20 sampled users, we also randomly sample up to 2 sessions labeled by our rule set as attacker-related and 1 session labeled as benign, where we define a \emph{session} to consist of all events with the same (IP address, user agent) pair value; all events within the same session are assigned the same label by our rule set. Across the sample of 20 users, we evaluate a total of 54 sessions, 34 of which are labeled as attacker-related sessions and 20 as benign.

\subsubsection{Establishing Ground Truth}
\label{sec:groundtruth}
In order to evaluate whether the labels our rule set applies to sessions are correct, we must develop a way to reason about what the ground truth labels for sessions are. Just knowing that a user has been compromised does not give us much information on which particular login events are performed by the attacker. In this section, we describe four basic indicators that we apply to each of the 54 sessions to help us gain confidence on what the ``true'' labels are for the sessions when evaluating the rule set. We note that since the four basic indicators discussed in this section are not perfect in terms of determining the true label for sessions, we also perform a more extensive manual analysis for sessions in which the basic indicators label a session as benign and our rule set labels the session as attacker-related to ensure comprehensiveness. Due to our conservative approach, we aim to limit the false positives of our rule set and thus obtain a more refined label through manual analysis when the basic indicators are not comprehensive enough. Throughout the remainder of this section, we refer to a compromised user out of our sample of 20 as Bob and one of Bob's sessions as $s$.
\newline\newline
\textbf{Phishing Email Indicator}. For some session $s$, we retrieve all emails sent from Bob's account within $\pm 5$ hours from a login event in $s$. The time window of $\pm 5$ hours serves as a heuristic for relating the email to the login event it is close in time to, as users may not send email immediately after logging into their accounts. In addition, there are sometimes delays as to when Office 365 creates login events in the data buckets for retrieval by Barracuda. 

Once all emails are retrieved that are close in time to $s$, we first determine if any of the emails were flagged as phishing by Barracuda; if so, then we assign the phishing email indicator for $s$ a value of 1. If none of the emails were flagged, we then iterate over all emails and manually label them as phishing or not, using properties of the email header and body. Our method for manually labeling emails as phishing is similar to approaches taken in previous work \cite{BECguard,ho2019detecting}, in which we first analyze the subject of an email in the context of the sender's organization and the number and domains of recipients of the email. For example, an email of the form "Bob has shared a file with you" sent to many recipients across many types of domains is very likely to be phishing. For emails in which the subject is not suspicious and the number of recipients is small, we look through the bodies of the emails along with any links present to determine if the domains of the links are unrelated to the sender's organization. For many of the emails that we looked at, these steps were sufficient to determine if emails were related to phishing or not. We assign the phishing email indicator for $s$ a value of 1 if there was at least one phishing email we labeled; else, we assign the phishing email indicator for $s$ a value of 0 if there were no flagged emails by Barracuda and no manually labeled phishing emails.
\newline\newline
\textbf{Inbox Rules Detection Indicator}. We retrieve all suspicious inbox and email forwarding rules detected by Barracuda that are $\pm 5$ hours from a login event during Bob's session $s$. An inbox rule detection indicates that a suspicious rule was created in a user's account, such as emails being forwarded to the trash or to an external account. The inbox rules detection indicator is assigned a value of 1 for session $s$ if at least one inbox rules detection exists close in time to $s$.
\newline\newline
\textbf{Interarrival Time Indicator}. The \textit{interarrival time} between 2 login events $e_1$ and $e_2$ is the absolute value of the difference in timestamps between $e_1$ and $e_2$. The general idea for including this interarrival time indicator is for detecting if the absolute time difference between login events for the same user with two different locations is shorter than the expected travel time between the two locations.
We first obtain the country subdivision that is most common in Bob's historical user profile (i.e. the country subdivision that is associated with the most number of Bob's historical login events). For simplicity, we call this country subdivision Bob's home territory. Then, for each of Bob's login events $e$ during session $s$, we compute the interarrival time between $e$ and the closest login event in time to $e$ that contains Bob's home territory. Among all events during Bob's session $s$, we take the smallest interarrival time and if that value is smaller than the expected travel time between Bob's home territory and the location mapped to session $s$, we mark the interarrival time indicator for $s$ with a value of 1. 
\begin{table}[t]
\caption{Illustration of an example of a set of login events, with country subdivision in the first column and timestamp in the second column.}
\vskip 0.15in
\begin{center}
\begin{small}
\begin{tabular}{lcccr}
\toprule
Country Subdivision & Timestamp & \\
\midrule
MO, US    & 2019-11-29 08:00:05 \\
IL, US    & 2019-11-29 15:06:45\\
IL, US    & 2019-11-29 21:14:32 \\
27, JP    & 2019-11-29 22:00:07\\
\bottomrule
\end{tabular}
\end{small}
\end{center}
\vskip -0.1in
\label{interarrival-table}
\end{table}

An example of a set of login events (anonymized for privacy) is shown in Table \ref{interarrival-table}. If IL, US is Bob's home territory and we are evaluating one of Bob's sessions tied to 27, JP, the interarrival time for this session would be about 46 minutes given that there was a login event from 27, JP 46 minutes after a login from IL, US. However, the expected travel time between Illinois and Japan is about 13 hours. Therefore, the session tied to 27, JP would be suspicious and would be marked with a value of 1 by the indicator. Note that in applying the indicator to some session $s$, we use Bob's home territory for computing interarrival times for login events during $s$ to reduce the amount of manual analysis needed to be done in evaluating the rule set (we had to manually look up the expected travel time between all sessions and the respective home territories for our random sample of 20 users). To make this indicator more general, for each event during $s$, we could calculate the smallest interarrival time between the event and any country subdivision within Bob's historical user profile. However, using the home territory for each user was sufficient for the evaluation. 
\newline\newline
\textbf{Tor Exit Node Indicator}. If the IP address for $s$ is a Tor exit node, then we assign the Tor exit node indicator for session $s$ as a 1.
\newline\newline
\textbf{Applying the Basic Indicators and Refinement}. For each of the 54 sessions across our random sample of 20 users that we evaluate our rule set on, we apply the four basic indicators described above. If at least one indicator labels a session with a value of \textbf{1}, then we say that it is an attacker-related session; otherwise, we label it a benign session. 

Out of the 54 sessions, there were seven that were labeled as benign by the basic indicators and attacker by our rule set. To ensure that we obtained the highest confidence ground truth label for these seven sessions (possible false positives), we performed manual analysis to obtain a more refined label. Each of the seven sessions involved a different compromised user. From their respective historical user profiles, four users primarily use US-based IP addresses and the remaining three primarily use IP addresses based in countries outside the US. Through our analysis that we present below for each of the seven sessions, we find that five of these sessions should be labeled as attacker-related by ground truth. For simplicity in our analysis, we will refer to each of the 7 sessions as session $x$, where $x$ is a number between 1--7. 

For session 1, the IP address mapped to a country that had never been seen before in the historical user profile. In addition, there was only one login event from this new country and it occurs implausibly soon after a benign login event where it is impossible to travel between the two locations within the interarrival time. The interarrival time indicator didn't flag for session 1 because interarrival times were only calculated with respect to user's home territories and not any location seen in a historical user profile. As a result, this session is truly an attack. In both session 2 and session 3, the user agent string matches that of the second randomly sampled session in which our rule set correctly labeled as attacker (verified via ground truth). As a result, we declare session 2 and session 3 as attacker-related sessions.

The analysis for session 4 and session 5 is very similar. Both sessions map to new countries that have never been seen before in their respective users' login events from August 1, 2019 - January 27, 2020. In addition, both sessions involve user agents that are totally different from what their respective users use during their benign sessions (historical login events + benign sessions labeled by rule set). In session 4, there are a total of 6 login events over a time period of 3.5 weeks from the new country. Halfway through the time period, there is an interspersed login mapped to the user's home territory (most common country subdivision in historical user profile). Then, 2 weeks later, there is a final login from the new country. The only way that session 4 could be benign is if the user decided to travel back-and-forth between the new country and their home territory over the 3.5 week window; based on the fact that there was one interspersed login from the user's home territory, the user would need to make 3 total back-and-forth trips between the home territory and new country over the 3.5 week period, which seems unlikely. Also, since the country has never been seen before throughout any of the user's previous login events, we declare this an attacker-related session. Similarly, in session 5, at least 4 back-and-forth trips between the home territory and the new country would be required over a 2 week period. As a result, session 5 is attacker-related.

Sessions 6 and 7 are likely benign sessions, as captured by the basic indicators above. For session 6, login events during the session do not happen close in time to other attacker sessions that our rule set correctly classifies for the user. In addition, the user agent doesn't stand out as suspicious and is a standard Firefox user agent string similar to the form "Mozilla/5.0 (Windows NT 10.0; Win64; x64) Gecko/20100101 Firefox/70.0". Even though the country mapped to session 6 is a new country that has never been seen before in the historical user profile, the user's organization has offices in this country. As a result, we believe this is a benign session. For session 7, the associated country subdivision has never seen before in the user's historical profile, but the country appears in all of the user's historical login events. This session was flagged by our rule set because the user agent had never been seen before in the historical user profile, as the the device that the user typically uses was of an older model. An example is if the user's historical login events frequently contain the user agent string \texttt{iPhone9C4} and this session in question contains the user agent string \texttt{iPhone10C2}. However, during the two-month window of login events that we applied our rule set on, the typical user updated their device before we see any session 7 events, but this update wasn't reflected in the historical user profile until after running the rule set on session 7. We also continue to see the use of session 7's user agent after the two-month window in login events tied to the user's home territory. As a result, we label session 7 as benign.

Therefore, through our manual analysis, we obtain more refined labels for the seven sessions mentioned above and find that five of the sessions are attacker-related and two are benign.

\begin{table}[t]
\caption{Evaluation results of our rule set. 'False Positives (FP)' shows the number of sessions that the rule set labels as attack but ground truth labels as benign. 'False Negatives (FN)' shows the number of sessions that the rule set labels as benign but ground truth labels as attack.}
\vskip 0.15in
\begin{center}
\begin{small}
\begin{tabular}{lcccr}
\toprule
Metric &  \\
\hline
Compromised Users & 20 \\
Sessions & 54 \\
\hline
False Positives (FP)   & 2 \\
False Positive Rate    & 11\% \\
Precision    & 94\% \\
\hline
False Negatives (FN)   & 9 \\
False Negative Rate    & 22\% \\
Recall    & 78\% \\
\bottomrule
\end{tabular}
\end{small}
\end{center}
\vskip -0.1in
\label{metric-table}
\end{table} 

\begin{table}[t]
\caption{Confusion Matrix for 54 sessions across 20 users.}
\setlength{\extrarowheight}{2pt}
\begin{tabular}{cc|c|c|c|}
  & \multicolumn{1}{c}{} & \multicolumn{3}{c}{True Label} \\
  & \multicolumn{1}{c}{} & \multicolumn{1}{c}{Attacker}  & \multicolumn{1}{c}{Benign}  & \multicolumn{1}{c}{Total} \\\cline{3-5}
            & Attacker & 32 & 2 & 34 \\ \cline{3-5}
Rule Set  & Benign & 9 & 11 & 20 \\\cline{3-5}
         Label & Total & 41 & 13 & 54 \\\cline{3-5}
\end{tabular}
\label{confusion-matrix}
\end{table}
\subsubsection{Evaluation Results}
\label{sec:rulesetresults}
Tables \ref{metric-table} and \ref{confusion-matrix} summarize the performance metrics of our rule set and display the confusion matrix for the 54 sessions we evaluate the rule set on across the set of 20 randomly sampled users. As mentioned above in Section \ref{sec:evaluationmethodology}, our rule set labeled 34 of the sessions as attacker and 20 of the sessions as benign. Based on the ground truth analysis discussed above, each session also has a ``true'' label. Our rule set generates 2 false positive sessions (FP) and a false positive rate of 11\%. \textit{Precision} is defined as the number of sessions that our rule set correctly marks as attacker divided by the total number of sessions that our rule set marks as attacker (true attacker sessions plus false positives). The precision for our rule set is 94\%. We base our evaluation numbers on ground truth labels that we assign to sessions and we acknowledge that these are not perfect. However, due to the extensive manual analysis we perform to obtain more refined labels after applying the four basic indicators as discussed in Section \ref{sec:groundtruth}, our source of ground truth is relatively comprehensive.

Our rule set also generates 9 false negative sessions (FN) and a false negative rate of 22\%. This seems to suggest that attackers show some level of sophistication in trying to evade detection (i.e. accessing user accounts with locations that blend or match with the user's typical login locations).

\section{Damage Prevention through Analysis of Phishing Emails}
\label{sec:damageprevention}
\noindent Out of the 159 compromised enterprise accounts that we analyzed, 11 had at least one email flagged as phishing during their attack windows. For each of the 11 enterprise accounts, we calculated the time difference between when the first phishing email was sent and when the first attacker login event occurred as classified by our rule set. 

\begin{figure}[t]
\begin{center}
\includegraphics[width=\textwidth]{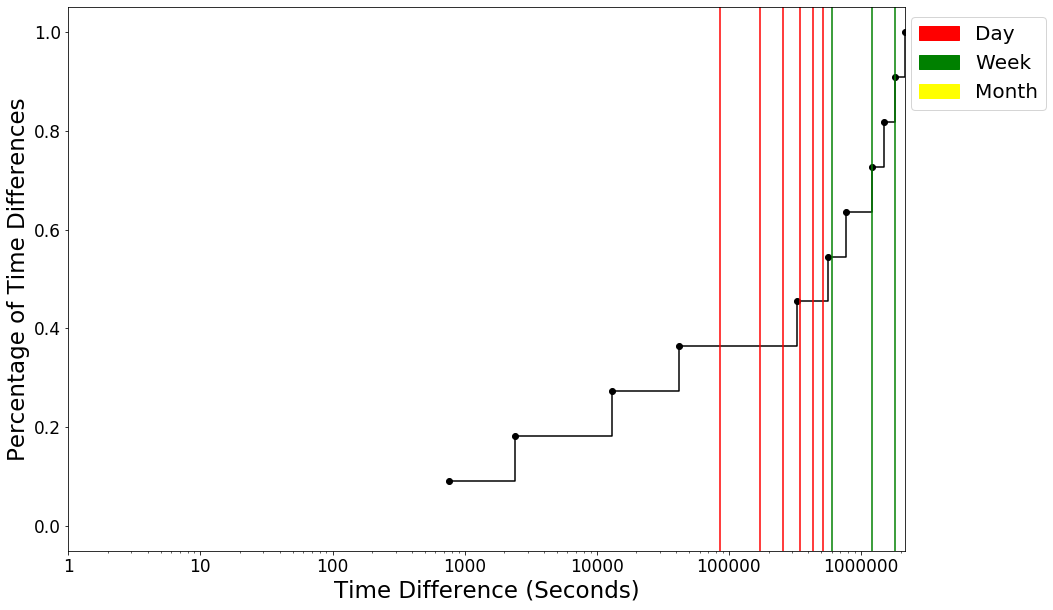}
\end{center}
\caption{\label{fig:firstloginphish} CDF of the distribution of differences in time in seconds between first phishing email and first attacker login event for each of the 11 compromised enterprise accounts that sent at least one phishing email during their attack windows. \textbf{Note}: The x-axis has been log-scaled.}
\end{figure}
Figure \ref{fig:firstloginphish} shows the distribution of time differences for each of the 11 compromised enterprise accounts. We can see that 4 out of 11 compromised accounts (37\%) had less than 1 day between the first phishing email and first attacker login event. The remaining 7 compromised accounts (63\%) had over 3 days of time difference.

\section{Modes of Attackers: Extended Analysis}
\subsection{Stability of IP Addresses}
\label{sec:stability}
One can argue that the low geolocation Jaccard similarity coefficients might be a result of attackers using unstable anonymized IP proxies or even Tor. For each of the 74 accounts that had max attacker interarrival times of more than 1 day, we computed the number of unique hours and number of unique country subdivisions seen across all attack events after the account's respective max attacker interarrival time. For each account, we calculated the following stability ratio of the form $$\text{stability} = \frac{\text{number of unique country subdivisions}}{\text{number of unique login hours}}.$$ If attackers are using unstable proxy services or Tor, we would expect this ratio to be large for many of the enterprise accounts. 
\begin{figure}[t]
\begin{center}
\includegraphics[width=\textwidth]{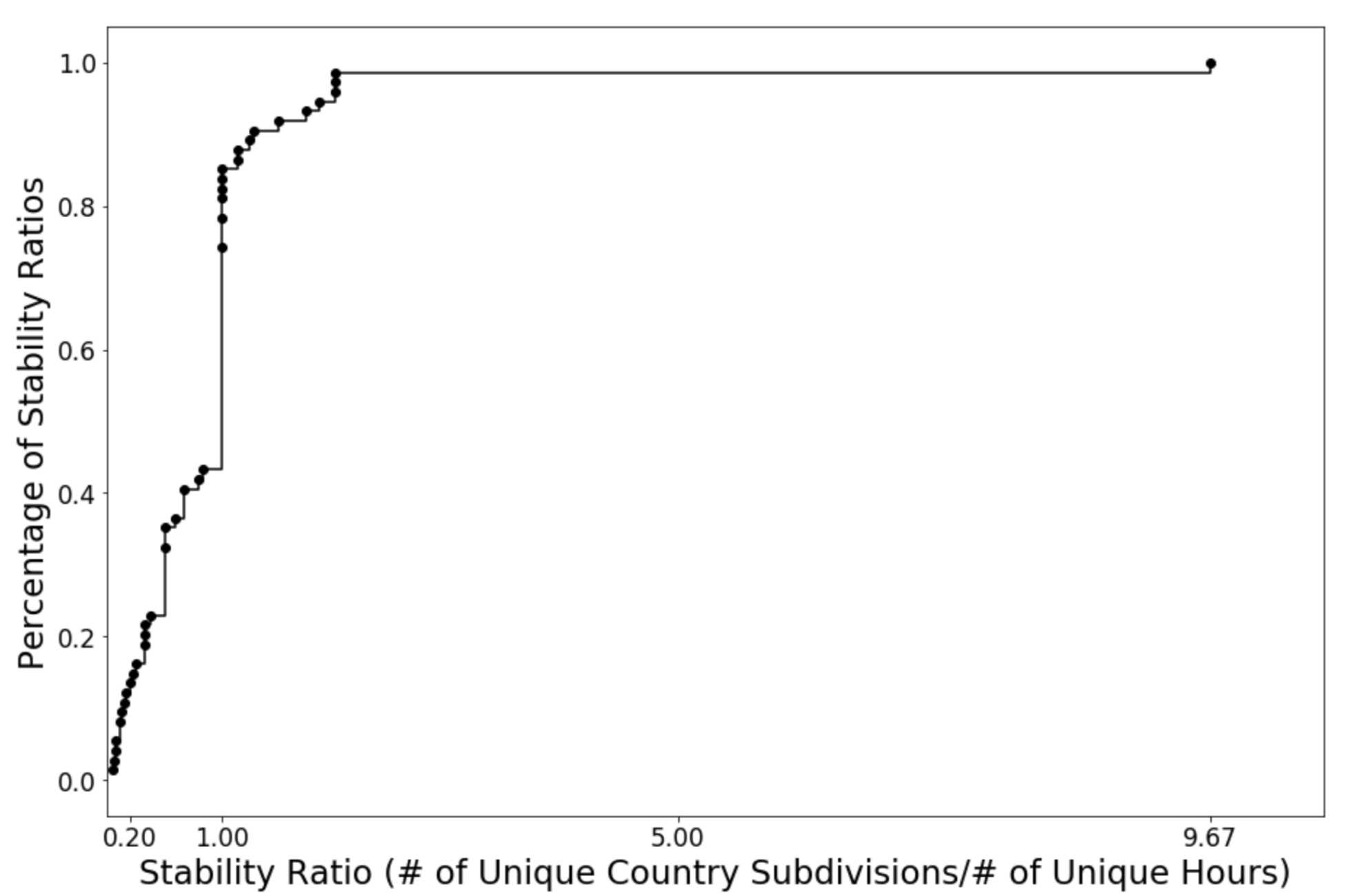}
\end{center}
\caption{\label{fig:stability} CDF of the Stability Ratios for each of the 74 Compromised Enterprise Accounts}
\end{figure}
As we can see from Figure \ref{fig:stability}, which shows a CDF of the stability ratios for each of the 74 enterprise accounts, 85\% of the accounts have stability ratios of at most 1 and 45\% of the accounts have stability ratios less than 1. After looking into the enterprise account that had a stability ratio of 9.67, it was obvious that the attacker was using a specialized proxy service that generated a different IP address upon each login. In general, if attackers are using proxy services to obtain IP addresses, these services seem to be fairly stable and as a result, geolocation seems to be a viable way to distinguish between different attackers.

\subsection{Activity Analysis of Specialized Attackers}
\label{sec:activitydiff}
For the 50 enterprise accounts in which we believe there are two sets of attackers (one set of attackers that performs the compromise and a second set of attackers that purchases the accounts and utilizes them), we are interested in determining if the second set of attackers inflicts more damage to the account than the first set. we developed an \textit{application access rate} metric that measures the number of Office 365 applications accessed by attack events divided by the number of unique hours the attack events span over a certain time period. For each of the 50 accounts, we computed the \textit{application access rate} before and after the max attacker interarrival time and in 30 of the 50 accounts (60\%), we find that the \textit{application access rate} after the max attacker interarrival time is larger than that before the max attacker interarrival time. Therefore, this analysis serves as a starting point for understanding the impact of credential selling and early mitigation of compromised enterprise accounts before they are sold to future attackers.

\begin{figure}[t]
\begin{center}
\includegraphics[width=\textwidth]{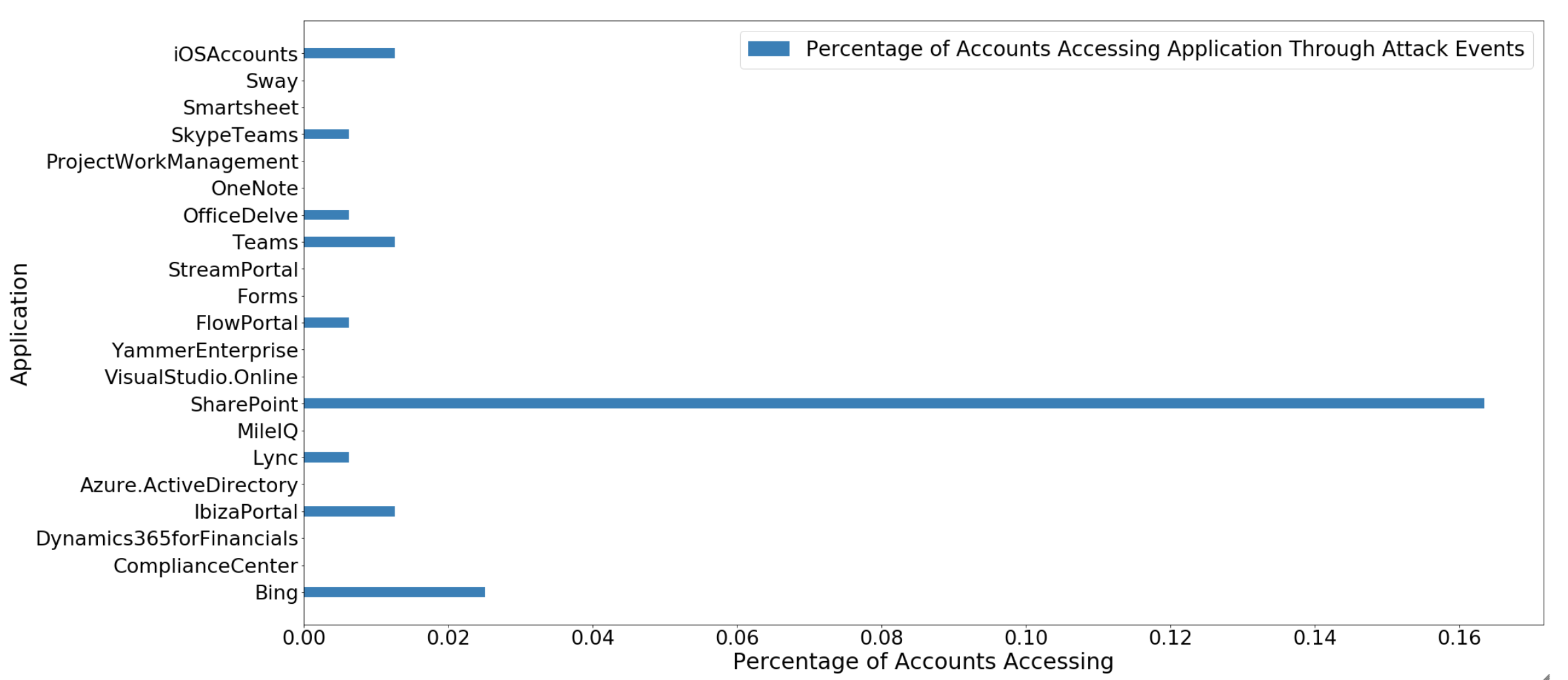}
\end{center}
\caption{\label{fig:applicationsfavor} Bar chart showing the percentage of enterprise accounts that access each non-email related Office 365 application through attack events.}
\end{figure}

\section{Applications that Attackers Favor}
\label{sec:appsfavor}
For each non-email-related Office 365 application that was accessed by at least one enterprise account in our dataset, we compute the percentage of accounts that access that application during their attacker events. The distribution is shown below in Figure \ref{fig:applicationsfavor}. 


\section{How Enterprise Accounts Are Compromised}
\label{sec:accountcompromises}
There are many ways in which enterprise accounts are compromised~\cite{companies:hacked}. Some common methods include phishing, lateral phishing~\cite{ho2019detecting}, password reuse, and the compromise of web-based databases. In this section, we analyze how enterprise accounts are compromised from the perspective of data breaches.

\begin{table}[t]
\caption{Table representing the number of organizations within each economic sector had at least one of its employee accounts found in the data breach.}
\vskip 0.15in
\begin{center}
\begin{small}
\begin{tabular}{lcccr}
\toprule
Economic Sector & Total \\
\hline
Consumer & 1 \\
Education   & 11\\
Food   &  1 \\
Government  & 1\\
Health & 2 \\
Industrials & 4\\
Technology   &  1 \\
Tourism  & 1\\
\hline
Grand Total & 23 \\
\bottomrule
\end{tabular}
\end{small}
\end{center}
\vskip -0.1in
\label{databreach-table}
\end{table}
\subsubsection{Data Breaches} We obtained data from a 3rd party data breach alert provider, whom we will keep anonymous for security purposes, that mines the criminal underground and dark web for compromised credentials involved in breaches of online company databases. From our dataset of 159 compromised enterprise accounts, 31 of the accounts (20\%) were involved in data breaches. Users of these accounts likely used their enterprise email address to create personal accounts on websites and when the websites' databases were breached, their associated personal account credentials were leaked. As a result, if these users reused credentials across their personal and enterprise accounts, their corresponding enterprise account was also likely compromised through the same data breach.

\begin{figure}[t]
\begin{center}
\includegraphics[width=\textwidth]{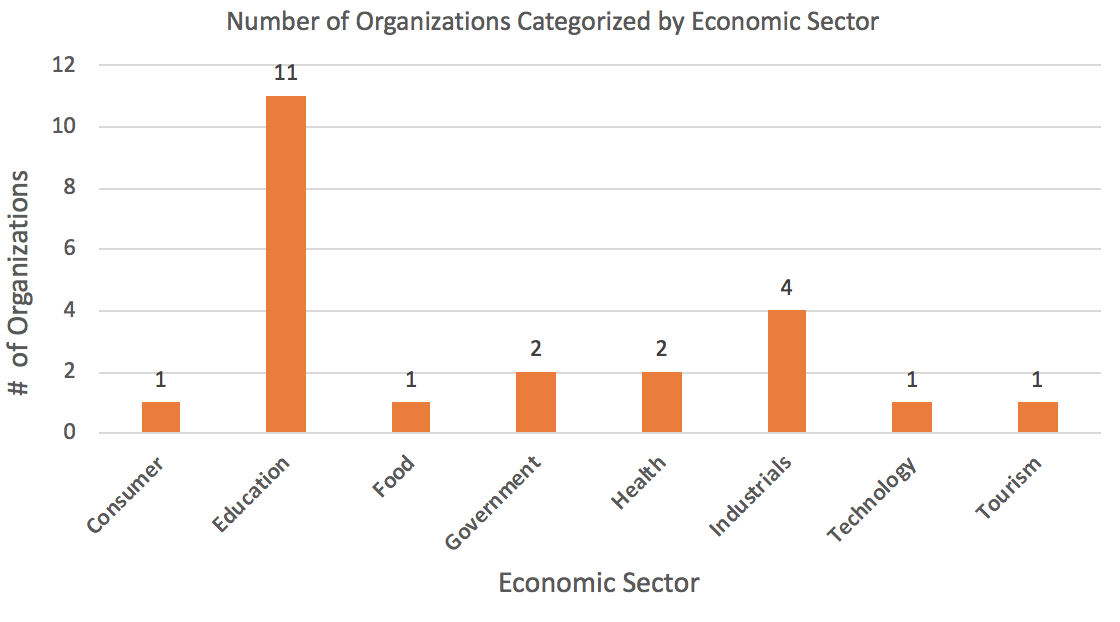}
\end{center}
\caption{\label{fig:spycloudbreaches} Bar chart of number of organizations within each economic sector that had at least one of its employee accounts found in a data breach.}
\end{figure}

Figure \ref{fig:spycloudbreaches} and Table \ref{databreach-table} display economic sectors and the number of organizations within those economic sectors that had at least one of their accounts involved in a data breach. The 31 enterprise accounts belong to 21\% of the organizations in our dataset (23 out of 111 organizations). We can see these 23 organizations span 8 of the 15 economic sectors. Although data breaches and credential leaks do not seem to discriminate against economic sectors, the education and industrials sectors seem to be hit the hardest in our dataset; there were 11 education organizations that had at least one compromised enterprise account found in a data breach and similarly, 4 industrials organizations. 

From our findings, educational accounts, such as those belonging to \texttt{.edu} organizations, are the most common accounts involved in data breaches and credential leaks. In many cases, users of these academic accounts tend to also create personal accounts on study websites and password reuse is common; as a result, if the databases backing the websites are breached, then the original academic accounts are also subject to compromise. There has been previous research in the field of analyzing the lure of compromising academic accounts, such as the work done by Zhang et al.~\cite{academic:compromise}. Zhang et al. note that academic accounts often offer free and unrestrained access to information due to less stringent security restrictions on these accounts. In addition, given that universities and schools are dormant for periods of time during the year and that upon graduation, users rarely access their educational accounts, attackers can go unnoticed for certain amounts of time in these accounts.

The findings in this section offer an insight into how enterprise accounts can be compromised. We saw that 21\% of enterprise accounts were found in a data breach of online company databases; although we don't know for sure if these enterprise accounts were compromised as a result of the data breach, we nevertheless show that data breaches are fairly common among enterprise accounts and credential reuse with personal accounts can cause a lot of damage. As a result, enterprises should frequently remind their employees of the dangers of credential reuse among their accounts to avoid additional compromises of their accounts.


\bibliographystyle{splncs04}
\bibliography{paper}
\end{document}